\documentclass[
]{style}

\usepackage{listings}
\usepackage[%
  bottom=3cm,  
  top=3cm,       
  left=3cm,
  right=3cm
]{geometry}
\usepackage{graphicx}
\usepackage{tcolorbox}
\usepackage{enumitem} 
\usepackage{multirow}
\usepackage{mdframed}
\usepackage{url}
\usepackage{hyperref}
\usepackage{booktabs}
\usepackage{array}
\usepackage{algorithm}
\usepackage{algpseudocode}
\usepackage{comment}
\usepackage{float}
\usepackage{tabularx}

\usepackage{pgfplots}
\pgfplotsset{compat=1.17} 

\begin{document}

\copyrightyear{2025}
\copyrightclause{Copyright for this paper by its authors.
  Use permitted under Creative Commons License Attribution 4.0
  International (CC BY 4.0).}


\title{Knowledge Base-Aware Orchestration: A Dynamic, Privacy-Preserving Method for Multi-Agent Systems}

\author{Danilo Trombino}[
orcid=0009-0000-6530-5081,
email=danilotrombino@gmail.com,
url=https://www.linkedin.com/in/danilotrombino/,
]
\fnmark[1]

\author{Vincenzo Pecorella}[%
orcid=0009-0006-6318-2843,
email=vincenzo.pecorella.vp@gmail.com,
url=https://www.linkedin.com/in/vincenzo-pecorella/,
]
\fnmark[1]

\author{Alessandro De Giulii}[%
email=alessandro.degiulii@gmail.com,
url=https://www.linkedin.com/in/degiulii/,
]
\fnmark[2]

\author{Davide Tresoldi}[%
email=d.tresoldi5@gmail.com,
url=https://www.linkedin.com/in/tresoldidavide/,
]
\fnmark[2]

\fntext[1]{These authors contributed equally to this work.}
\fntext[2]{These authors contributed equally as advisors.}
\begin{abstract}
Multi-agent systems (MAS) are increasingly tasked with solving complex, knowledge-intensive problems where effective agent orchestration is critical. Conventional orchestration methods rely on static agent descriptions, which often become outdated or incomplete. This limitation leads to inefficient task routing, particularly in dynamic environments where agent capabilities continuously evolve. We introduce \textit{Knowledge Base-Aware (KBA) Orchestration}, a novel approach that augments static descriptions with dynamic, privacy-preserving relevance signals derived from each agent’s internal knowledge base (KB). In the proposed framework, when static descriptions are insufficient for a clear routing decision, the orchestrator prompts the subagents in parallel. Each agent then assesses the task's relevance against its private KB, returning a lightweight  ACK signal without exposing the underlying data. These collected signals populate a shared semantic cache, providing dynamic indicators of agent suitability for future queries. By combining this novel mechanism with static descriptions, our method achieves more accurate and adaptive task routing preserving agent autonomy and data confidentiality. Benchmarks show that our KBA Orchestration significantly outperforms static description-driven methods in routing precision and overall system efficiency, making it suitable for large-scale systems that require higher accuracy than standard description-driven routing.
\end{abstract}


\maketitle

\section{Introduction}

Multi-Agent Systems (MAS) are increasingly deployed to tackle complex, cross-domain problems in areas such as autonomous robotics, smart city infrastructure, collaborative software development, distributed scientific research, and advanced financial modeling \cite{smythos2025multiagent}. Their strength lies in decomposing large problems into smaller, specialized tasks handled by autonomous agents working in concert.  

Central to unlocking this potential is \emph{agent orchestration},  the process of determining which agent or set of agents should handle a given task. Whether implemented via a centralised controller or a distributed invocation protocol, orchestration directly impacts the overall efficiency, accuracy, and adaptability of the system \cite{arxiv2025llmcollab}.  

However, the prevailing paradigm for orchestration relies on matching tasks to agents based on \emph{static, predefined descriptions} of their capabilities. This approach is fundamentally brittle, especially in knowledge-intensive environments. For instance, an agent specialized in data analysis might acquire new insights about a specific machine learning model through its operations, but its static profile would not reflect this updated expertise. Consequently, the orchestrator, blind to this evolution, may misroute a relevant task to a less qualified agent, leading to suboptimal outcomes, wasted resources, and a system that cannot adapt.

This paper directly addresses these shortcomings by proposing \textbf{Knowledge Base-Aware (KBA) Orchestration}, a framework that moves beyond rigid profiles by incorporating dynamic, privacy-preserving signals from the agents themselves. Our work makes the following contributions:
\begin{itemize}
\item We introduce a novel orchestration mechanism where agents provide real-time relevance scores based on their private knowledge bases, enhancing routing decisions without compromising confidentiality.
\item We present the design of a shared semantic cache that allows the orchestrator to efficiently access and leverage these dynamic signals for context-aware task assignments.
\item We provide a comprehensive empirical evaluation showing that our KBA approach significantly outperforms traditional static methods in both routing precision and overall system efficiency.
\end{itemize}

\section{Background and Related Work}

Numerous efforts have been made to develop Multi-Agent Systems (MAS), yet no single architecture has emerged as universally superior. Major technology companies have proposed various frameworks and architectures, each tailored to specific use cases and application domains. Given that ``AI agents are generally defined as a class of interactive systems that can perceive visual stimuli, language input, and other environmental data, and can produce meaningful embodied actions'' \cite{durante2024agentaisurveyinghorizons}, a wide range of architectural and methodological approaches have been explored to enable effective agent interaction.

In the following sections, we will analyze orchestration paradigms within the context of a prevalent architecture: the \textbf{centralised Multi-Agent System}. In this model, all user queries are funneled through a single orchestrator that transparently selects the appropriate agent.

This centralised model offers several advantages: It simplifies the user experience, enforces consistent request handling, enables global logging and monitoring, and allows the system to apply advanced routing logic (e.g. intent classification, load balancing, or escalation rules) without requiring changes to individual agents. 
Notable examples of this architecture include the pattern \emph{Coordinator/Dispatcher} in Google's Agent Development Kit (ADK)~\cite{googleADKMultiAgents}, which formalizes the selection and delegation process among multiple agents, and emerging corporate AI suites such as Microsoft Copilot~\cite{microsoftCopilot} and Google AgentSpace~\cite{googleAgentSpace}. These suites are currently centralizing access to multiple agents through a single unified interface, but in most cases, the choice of which agent to use still requires manual selection or configuration by the user.

However, the currently-adopted orchestration mechanisms that will be discussed in section~\ref{sec:agent-orchestration} face significant challenges when applied in this centralised multi-agent orchestration setting. As will be shown in the following discussion, the two approaches presented encounter limitations that reduce their effectiveness in this context.

\subsection{Orchestration Mechanisms Among Agents in Multi-Agent Systems and their limitations}
\label{sec:agent-orchestration}

One of the core aspects of a MAS lies in its \textbf{orchestration mechanisms}. This aspect concerns how agents are coordinated, activated, and interconnected to achieve collective goals, as examined in the following section.

A robust MAS requires a well-defined orchestration protocol to engage the appropriate agents. These mechanisms typically fall into two categories: \textit{Deterministic} and \textit{Description-Driven} Orchestration , each with significant trade-offs in a centralised, dynamic setting.

\subsubsection{Deterministic Orchestration} A structured, predictable approach where agents are activated according to a predefined control flow. The orchestration logic is fully specified in advance, with no runtime computation or inference to determine which agents to call. Whether agents execute in a linear pipeline or concurrently in distinct branches, the orchestration plan is hard-coded.

However, this rigid method is unsuitable for a multi-purpose system where user requests target unpredictable domains. A predefined workflow cannot dynamically adapt to diverse queries. Invoking all agents for every request is computationally wasteful and inefficient, while invoking a fixed subset is guaranteed to fail for out-of-scope tasks. This leads to degraded performance, especially in high-load, cost-sensitive environments.

Examples of this approach include simple Sequential Workflows or more complex layered architectures like the Mixture of Agents model ~\cite{wang2024mixtureofagentsenhanceslargelanguage}: 
\begin{itemize}
    \item \textbf{Sequential workflow}: Agents process messages in a fixed order. Each agent receives the predecessor’s output, performs its task, and passes the result onward, producing a deterministic, reproducible pipeline.
    \item \textbf{Mixture of agents}: Introduced by Wang et al.~\cite{wang2024mixtureofagentsenhanceslargelanguage} and extended in Microsoft’s AutoGen framework~\cite{microsoftAutoGenMixtureOfAgents}, this architecture defines a layered hierarchy inspired by feedforward neural networks. The input task is distributed to the first layer of agents, an \textit{aggregator agent} collects their responses, and forwards the updated information to the subsequent layer. At each stage, \textit{all nodes} are activated, and the outputs of agents in the $n$-th layer are aggregated by an \textit{aggregator agent} positioned between the $n$-th and $(n+1)$-th layers.

\end{itemize}

\begin{figure}[h]
  \centering
  \includegraphics[width=0.9\textwidth]{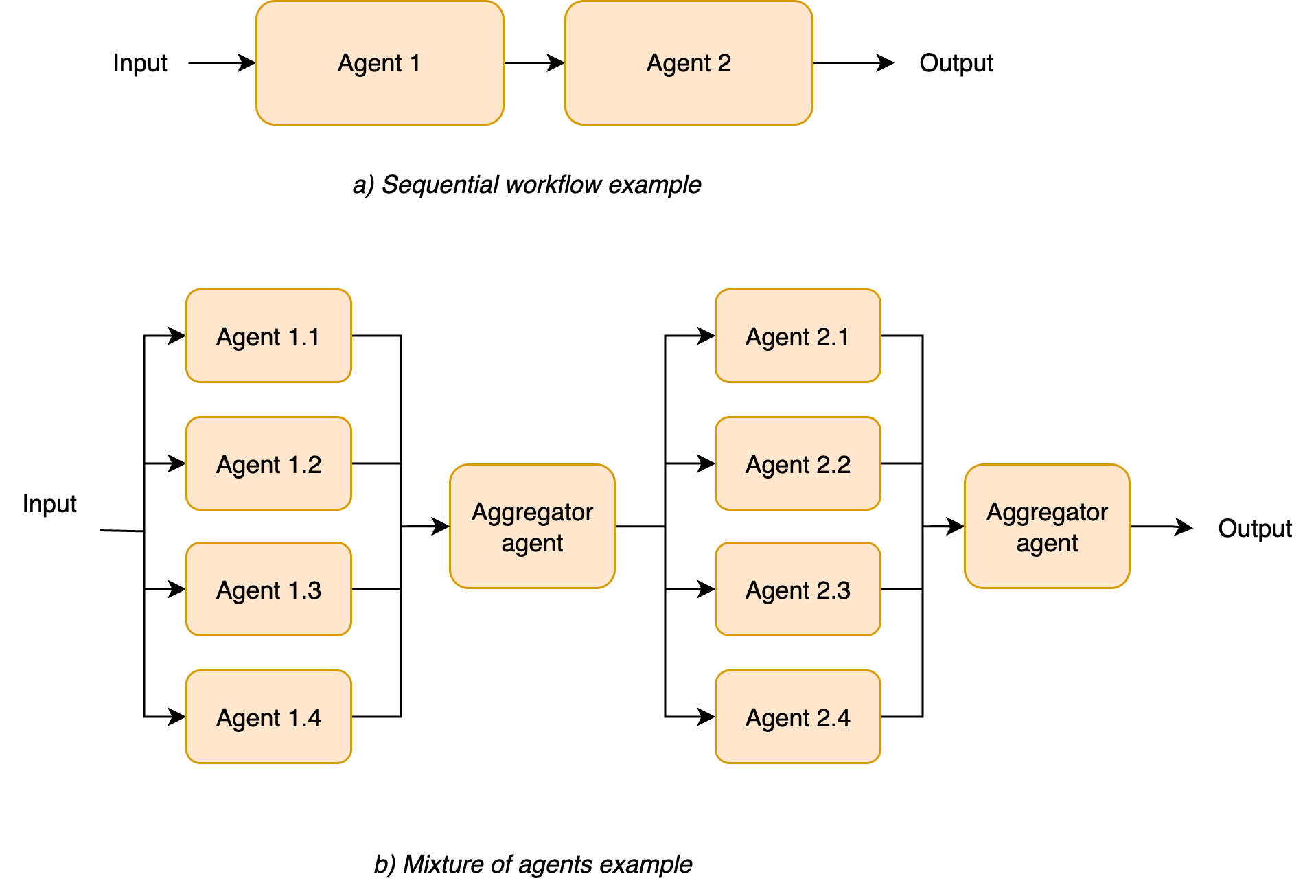}
  \caption{Examples of deterministic orchestration mechanisms}
  \label{fig:deterministic_invoke}
\end{figure}

\subsubsection{Description-Driven Orchestration}
A more flexible approach is represented by the description-driven orchestration. Here, each agent publishes a textual summary of its capabilities, the \textit{agent card}. At runtime, an orchestrator, typically using an LLM, matches the task's intent against these descriptions to select the most suitable agent.

\begin{figure}[h]
  \centering
  \includegraphics[width=0.9\textwidth]{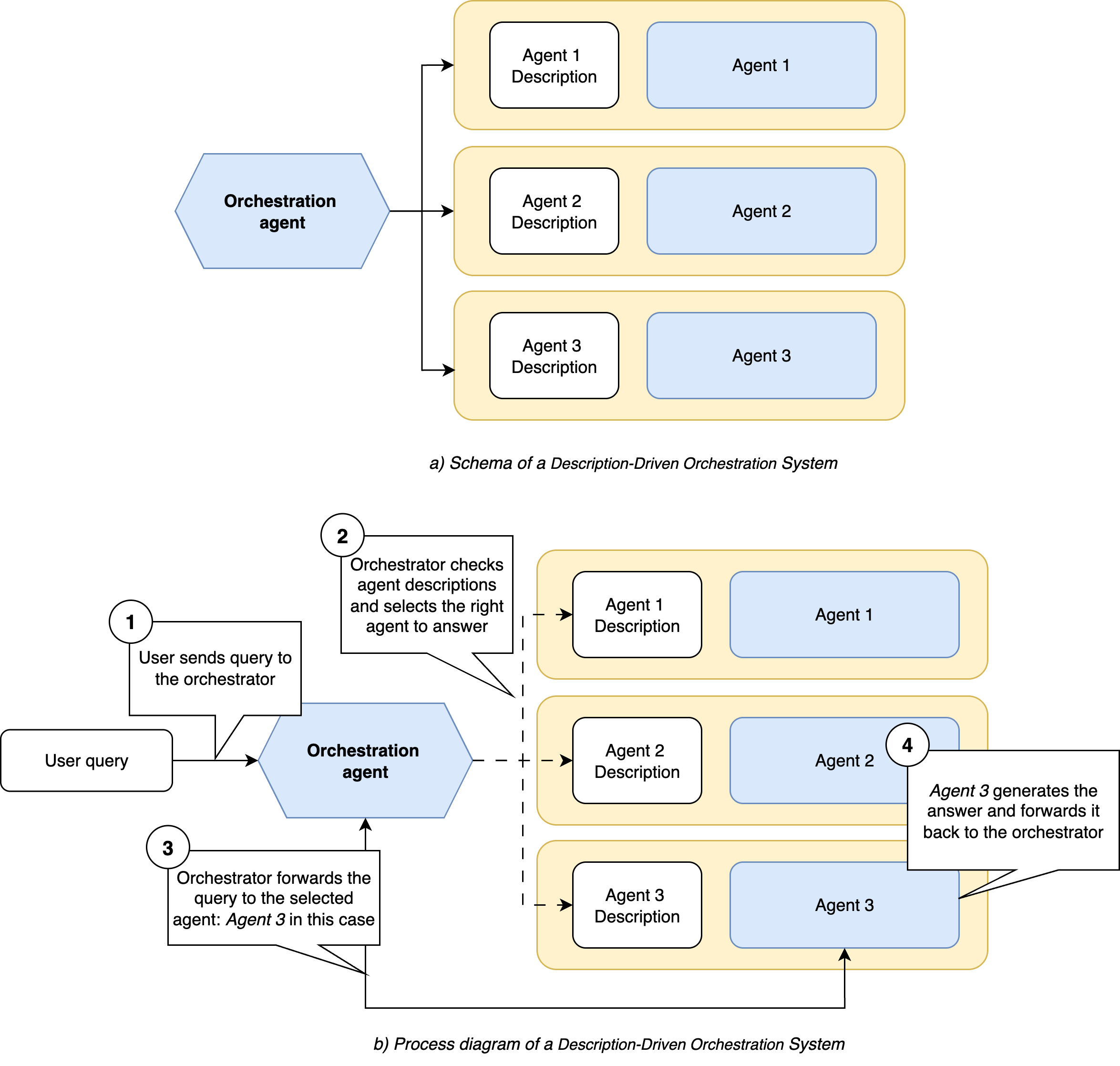}
  \caption{%
        Illustration of a \emph{Description-Driven Orchestration System}. 
        (a) High-level architecture showing the orchestrator and its connected agents with capability descriptions. 
        (b) Example process flow for description-driven orchestration within the centralised system, from user query to agent selection and response delivery.
    }
  \label{fig:description_based_invoke}
\end{figure}

A common implementation is the \emph{Coordinator/Dispatcher} model in Google’s Agent Development Kit (ADK)~\cite{googleADKMultiAgents}, also used in AWS’s Multi-Agent Orchestrator~\cite{awsMultiAgentOrchestrator}. Related techniques include LLM-driven selection or use of agent cards~\cite{google_a2a_docs_agent_card}.

Although common in multi-domain systems, this method has three major drawbacks:
\begin{itemize}
    \item \textbf{Incomplete task–agent alignment}: Descriptions may not capture all relevant capabilities, leading to misrouting, particularly in PaaS-style multitenant setups with generic or bad optimized descriptions.
    \item \textbf{Semantic overlap and ambiguity}: For reliable LLM-based selection, descriptions should be semantically distinct. In practice, overlaps cause confusion and incorrect routing.
    \item \textbf{Resource inefficiency}: Injecting all agent descriptions into every LLM prompt increases cost, latency, and confusion risk, especially with overlapping or redundant entries.
\end{itemize}

\paragraph{Example: routing ambiguity from vague agent cards}

Consider a system with a central orchestrator and two agents with the following agent cards:

\begin{tcolorbox}[title=Example of a Multi-Agent system, colback=gray!5!white, colframe=gray!75!black]
\textbf{Office management agent:} ``Handles issues related to office infrastructure, such as desks, monitors, seating arrangements, meeting rooms, and office spaces.''

\textbf{Tech support agent:} ``Solves problems concerning IT infrastructure, including computer malfunctions, application errors, and issues with company systems.''
\tcblower
\medskip
\textit{\textbf{User query:} ``My badge doesn’t work anymore, what should I do?''}
\end{tcolorbox}

In many real-world organizations, physical access badges fall under the responsibility of office management or facilities teams. In the example, the correct agent would therefore be the \textit{Office management agent}.  
However, the orchestrator has no explicit information in the agent descriptions about badges, building access, or similar terms. This means that when it tries to match the user’s request to an agent, it finds no clear evidence pointing to the right one.

The problem is compounded by indirect cues:  
\begin{itemize}
    \item The phrase ``doesn't work'' might be interpreted as a technical failure, which aligns more closely with the Tech Support’s description of ``malfunctions'' or ``errors.''
    \item Both descriptions use broad terms like ``infrastructure'' and ``systems,'' which can cover many areas and create semantic overlap.
\end{itemize}

Because the orchestrator lacks a decisive signal, it is forced to make a guess. In practice, this often results in it selecting the Tech support agent, not because it is the right choice, but because the language of the request sounds vaguely technical.  
This demonstrates a key weakness of pure description-driven invocation: when important responsibilities are not clearly expressed in the descriptions, the system cannot reliably determine the correct routing. Instead, it chooses based on partial or misleading signals, leading to incorrect assignments.

\subsection{Naive Description Expansion: A short-term fix with long-term risks}

A straightforward way to reduce misrouting is to expand each agent’s description with exhaustive lists of examples (e.g., explicitly adding ``badge access'' under the Office management agent). While this can resolve individual gaps, the approach is fundamentally unscalable. 

Attempting to anticipate every possible query would require bloating the descriptions until the orchestrator holds a near-complete copy of each agent’s knowledge. This effectively centralizes domain expertise at the orchestration layer, defeating the purpose of a modular, specialized multi-agent architecture.

Such centralization not only erodes the benefits of separation of concerns and independent agent evolution, but also introduces significant security risks. Sensitive or privileged information that should remain compartmentalized within an agent would instead be embedded in the orchestrator’s prompt context, increasing the surface area for potential data leakage. Instead, a viable solution must improve routing accuracy without centralizing knowledge or compromising the principles that make a multi-agent system effective.

\section{Methodology}
\label{section:methodology}

Our Knowledge Base-Aware (KBA) Orchestration overcomes the limitations of static routing by introducing a dynamic, multi-stage orchestration architecture. While traditional systems rely on a single classification step using agent descriptions (often known as \emph{agent cards}), our approach enhances this by delegating to subagents the verification within their private Knowledge bases. In this phase agents are asked to return an ACK that is used by the orchestrator to redirect the user's request to the most appropriate agent.

\subsection{Orchestration Flow}

\begin{figure}[h]
    \centering
    \includegraphics[width=0.9\linewidth]{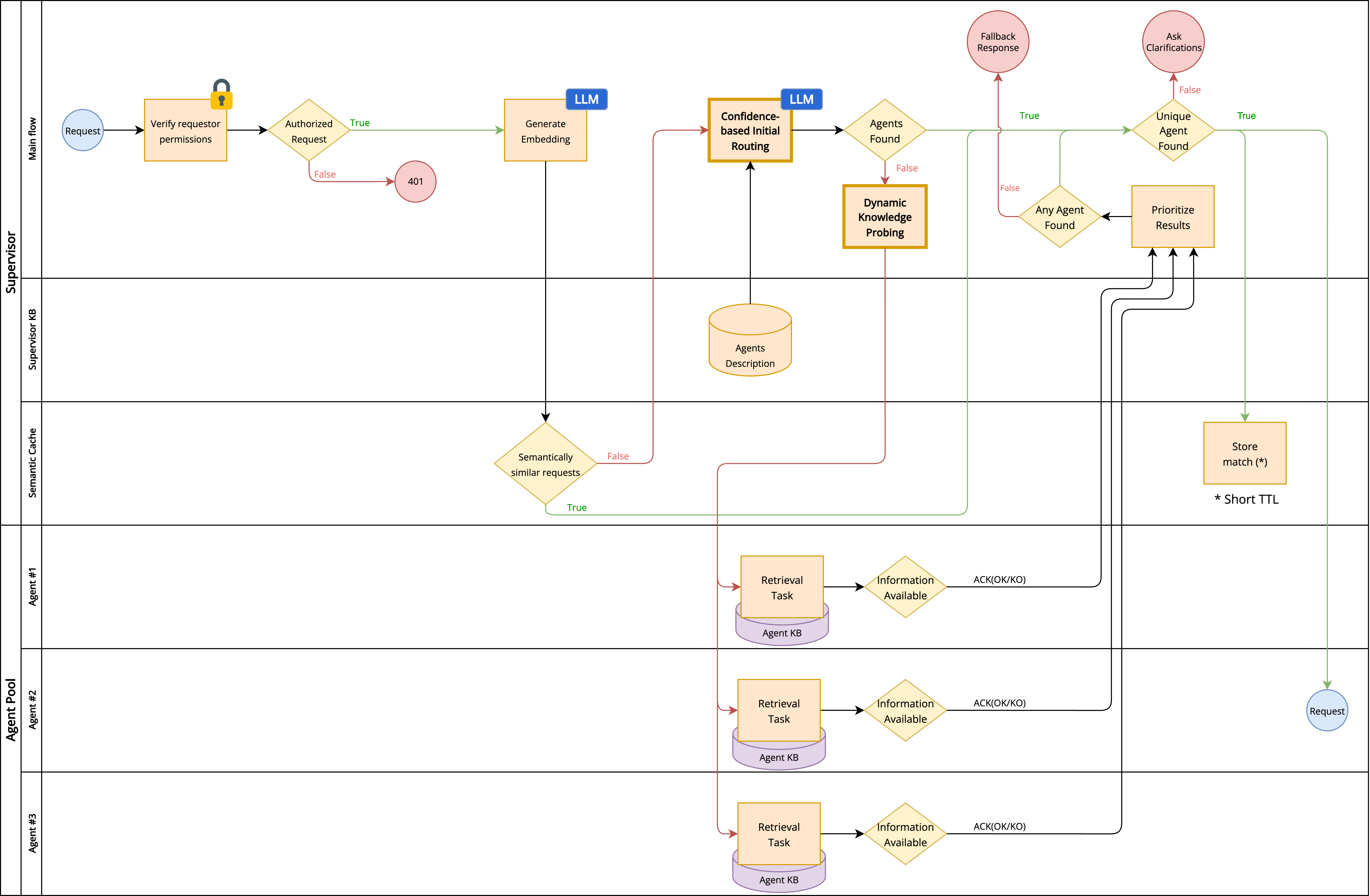}
    \caption{The KBA Orchestration Flow. A request first checks the semantic cache. On a miss, the orchestrator performs an initial classification. Low-confidence results trigger Dynamic Knowledge Probing. The final, validated result is then stored in the cache.}
    \label{fig:flow}
\end{figure}

Our \emph{Knowledge-Aware Orchestrator} introduces three key components to traditional orchestration approaches:  
(1) a Confidence-based Initial Router,  
(2) a Dynamic Knowledge Probing mechanism, and  
(3) a Semantic Cache.  
The interaction between these components is shown in Figure~\ref{fig:flow}.
The process evaluates the fastest path first (the cache) and only engages more complex components when necessary. The flow consists of the following ordered steps.

\subsubsection{Semantic Cache Lookup}
Upon receiving a query, the orchestrator first checks the semantic cache. If a semantically similar query has been successfully routed recently, the decision is reused instantly, providing the fastest possible response. To reduce repeated probing costs, this key-value cache stores successful routing decisions keyed by semantic embeddings of the queries. Cache entries expire according to the mechanisms outlined in \ref{subsection:cache_invalidation}.

\subsubsection{Confidence-based Initial Routing}
On a cache miss, the orchestrator performs an initial routing attempt. It prompts the LLM to assign confidence scores to each agent based on their static descriptions. The prompt is carefully constructed to instruct the LLM to signal ambiguity by providing a low confidence score. If the top score exceeds a configurable threshold ($\tau$), the query is routed directly to that agent. Otherwise, the routing is considered uncertain and escalated to the \textit{probing tool}. Within the prompt of the orchestrator the \textit{probing tool} is described as the authoritative source in ambiguous cases, ensuring correctness even at higher computational cost, while still enabling efficient direct routing when confidence is high.

\subsubsection{Dynamic Knowledge Probing}
\begin{figure}[h]
    \centering
    \includegraphics[width=0.9\linewidth]{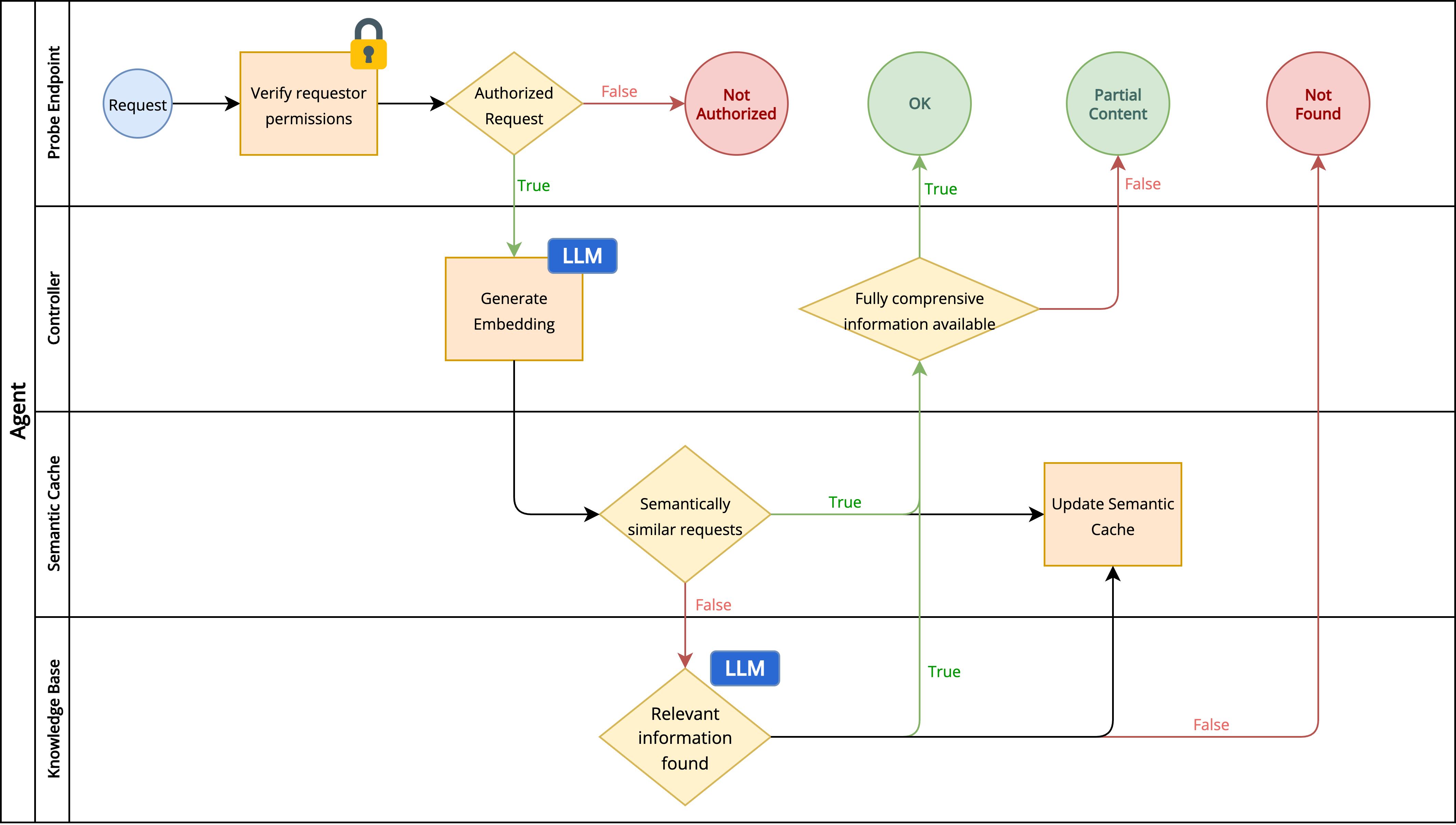}
    \caption{Agent-Side Advanced Probing Flow. Agents check permissions, consult the semantic cache, and query the knowledge base, returning \emph{OK}, \emph{Partial Content}, or \emph{KO} accordingly.}
    \label{fig:advanced-probing}
\end{figure}

If the initial confidence is below $\tau$, the orchestrator initiates the probing stage. It issues a parallel query to all candidate agents, asking each to verify internally if they can handle the request. Agents respond with a simple, lightweight acknowledgment (e.g., OK/KO) based on a search of their private knowledge base (the retrieval phase), without exposing any content. This binary mechanism can be extended to support more granular responses like \textit{Partial Content} or a confidence score. If only one agent responds positively, the query is routed there. If multiple respond, the user may be prompted to choose. Figure~\ref{fig:advanced-probing} illustrates the agent-side logic, which includes permission checks, semantic cache lookups, and knowledge-base searches. 

The framework treats each candidate agent's retrieval phase as a \emph{black box} from the orchestrator's perspective. This design enables maximum flexibility along two axes:
\begin{itemize}
    \item The nature of the underlying knowledge
    \item Requirements on latency, efficiency, and accuracy
\end{itemize}
Because the optimal choice is context-dependent and left to the system designer, we outline several representative implementation patterns below.

\paragraph{Traditional knowledge bases}
For document-centric knowledge bases (e.g., wikis, manuals, policy docs), the retrieval phase can be implemented as similarity search in an embedding space. Given a user query $q$, retrieve the top-$k$ most similar fragments $\{f_i\}$ (either pre-computed summaries or chunked excerpts). Let $s^\star=\max_i \mathrm{sim}(e(q), e(f_i))$. The lightweight acknowledgment is derived from $s^\star$: if $s^\star \ge \theta$ (a calibrated threshold), the agent returns \texttt{OK}; otherwise \texttt{KO}. Generation of the final, user-facing answer is deferred and executed only if the orchestrator selects this agent as the main responder.

\paragraph{MCP-based agents}
For agents implemented within the MCP framework, the retrieval phase can be realized as a binary (or graded) LLM classifier that conditions on the user query and the MCP server’s available \emph{Tools}, \emph{Resources}, and \emph{Prompts}. The classifier returns \texttt{OK}/\texttt{KO} based on whether the question can be addressed via at least one tool or by the content of the resources.

\paragraph{Multilayered agents}
If agents are organized hierarchically (i.e., beyond a simple two-level orchestrator–agent design), the same probing and retrieval principles can be applied recursively. Each sub-agent performs permission checks, semantic cache lookups, and knowledge-base searches, returning an \texttt{OK}/\texttt{KO} (optionally with confidence). A parent node aggregates these signals and responds upstream with its own lightweight acknowledgment.

\subsubsection{Cache Population}
Once a definitive route is determined (either through high-confidence initial routing or knowledge probing), the orchestrator stores the query's semantic embedding and the successful agent choice in the semantic cache. This ensures that the knowledge gained from this interaction is used to accelerate future, similar requests. 

\subsubsection{Cache Invalidation}
\label{subsection:cache_invalidation}
Semantic cache invalidation represents a fundamental paradigm shift from traditional exact-match caching to similarity-based approximate caching in high-dimensional vector spaces. Unlike conventional caching systems that invalidate entries based on exact key matches or time-based policies, semantic caching systems must navigate the complex geometry of embedding spaces where invalidation creates hyperspherical "holes" that affect semantically related cached entries.

\paragraph{The geometric nature of semantic invalidation}
In traditional caching systems, invalidation is a discrete operation: a cache entry either exists or it doesn't. Semantic caching, however, operates in continuous vector spaces where invalidation affects not just the target entry but also semantically related entries within a defined similarity radius. This creates what we term \textit{invalidation spheres}: hyperspherical regions in the embedding space where cached content is considered stale or incorrect.
The invalidation process in our KBA system follows a three-step  procedure:

\begin{enumerate}
    \item \textbf{Embedding generation}: When an invalidation request is received (e.g., when agent knowledge has been updated), the system first generates an embedding vector $\mathbf{v}_{inv}$ for the topic or content area to be invalidated using the same embedding model employed for cache storage.

    \item \textbf{Similarity search and threshold determination}: The system performs a vector similarity search within the cache to identify all entries within a defined similarity threshold $\theta_{inv}$ of the invalidation embedding. This threshold represents the radius of the invalidation hypersphere and is a critical parameter that determines the scope of invalidation.

    \item \textbf{Sphere-based deletion}: All cached entries $\mathbf{v}_i$ satisfying $\text{sim}(\mathbf{v}_{inv}, \mathbf{v}_i) \geq \theta_{inv}$ are removed from the cache, effectively creating a ``hole'' in the vector space representation.
\end{enumerate}

\paragraph{Mathematical formalization of invalidation boundaries}

The invalidation region can be formally defined as a hypersphere $S(\mathbf{c}, r)$ in the $d$-dimensional embedding space:

\begin{equation}
S(\mathbf{c}, r) = \{\mathbf{x} \in \mathbb{R}^d : ||\mathbf{x} - \mathbf{c}||_2 \leq r\}
\end{equation}

where $\mathbf{c}$ represents the centroid of the invalidation topic and $r$ is the invalidation radius derived from the similarity threshold. For cosine similarity, this relationship becomes:

\begin{equation}
r = \sqrt{2(1 - \theta_{inv})}
\end{equation}

This geometric interpretation reveals why semantic invalidation is fundamentally different from traditional approaches: rather than removing discrete entries, we are carving out continuous regions of the semantic space.

\begin{figure}[h]
    \centering
    \includegraphics[width=0.9\linewidth]{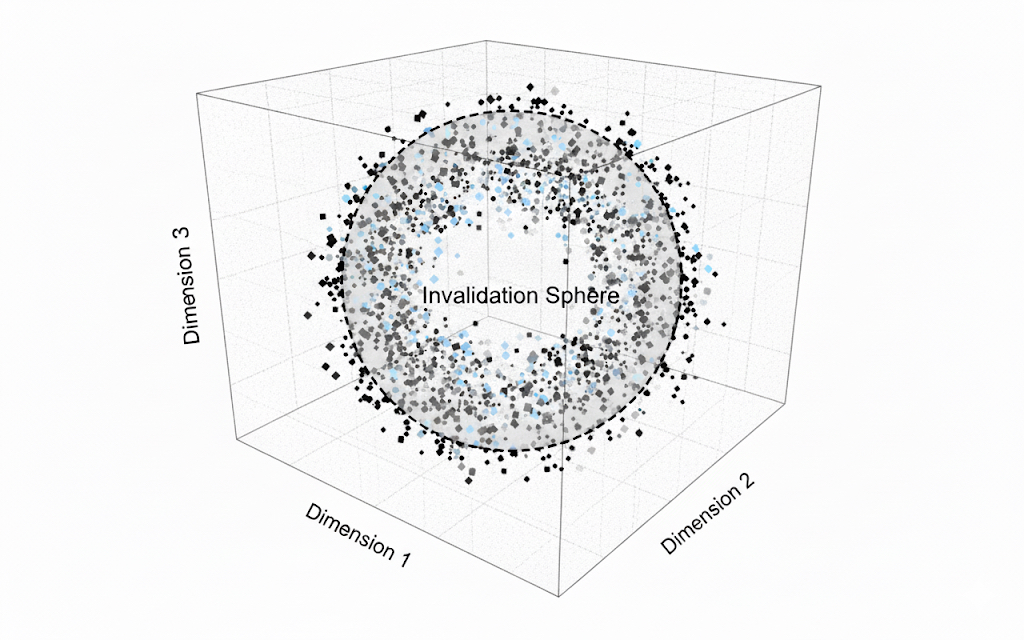}
    \caption{Cache invalidation through vector similarity can be visualizes as a hole within the embedding vector space}
    \label{fig:cache_invalidation}
\end{figure}

\paragraph{The threshold optimization challenge}

Determining the optimal invalidation threshold $\theta_{inv}$ a priori represents one of the most significant challenges in semantic cache invalidation. This threshold must balance two competing objectives:
\textit{precision} vs \textit{coverage} trade-off.

A high threshold (e.g. 0.95 cosine similarity) ensures that only highly similar content is invalidated, minimizing false positives but potentially leaving stale content in the cache. Conversely, a low threshold (e.g. 0.7) provides broader coverage but may invalidate semantically distinct but related content unnecessarily.

Based on current research and production deployments, we propose a \textit{dynamic threshold adaptation strategy} that considers multiple factors:

\begin{enumerate}
    \item \textbf{Domain specificity}: Technical domains requiring high precision (legal, medical) should employ higher thresholds (0.9-0.95), while general-purpose applications can operate effectively with moderate thresholds (0.8-0.85).
    
    \item \textbf{Cache staleness tolerance}: Time-sensitive information requires aggressive invalidation with lower thresholds, while stable knowledge bases can tolerate higher thresholds.
    
    \item \textbf{Historical performance metrics}: The system can learn optimal thresholds by monitoring false positive and false negative rates over time, adjusting thresholds based on observed cache hit accuracy.
\end{enumerate}

\subsection{Algorithmic Implementation}

To formalize the orchestration flow, we present the core logic in two algorithms. Algorithm~\ref{alg:route-request} details the orchestrator's end-to-end process for routing a request, encompassing the semantic cache lookup, the confidence-based initial attempt, and the dynamic knowledge probing fallback.

For the system to operate correctly, each participating agent must expose a compatible interface. Algorithm~\ref{alg:agent-side-retrieval-request} specifies this required agent-side retrieval logic, defining how an agent must process a probe from the orchestrator and return a valid response.

\begin{algorithm}[!h]
\small  
\caption{Knowledge Base-Aware Orchestration flow}
\label{alg:route-request}
\begin{algorithmic}[1]
\Function{RouteRequest}{Query $Q$, AgentPool $A$, Threshold $\tau$}
    \State $E \gets$ Embed($Q$)
    \State $C \gets$ SemanticCache.FindSimilar($E$)
    \If{$C \neq \varnothing$}
        \State \Return Handoff($C$, $Q$)
    \EndIf
    \State $(Best, UseProbing) \gets$ LLM.Classify($Q$, $A.GetAgentCards()$, $\tau$)
    \If{UseProbing}
        \State $R \gets$ KBAwareService.ParallelProbe($Q$, $A$)
        \State $Capable \gets$ \{ $a_i \in A \mid R[i] = \text{TRUE}$ \}
        \If{$|Capable| = 1$}
            \State SemanticCache.Store($E$, $Capable[0]$)
            \State \Return Handoff($Capable[0]$, $Q$)
        \ElsIf{$|Capable| > 1$}
            \State $F \gets$ ResolveAmbiguity($Capable$)
            \State SemanticCache.Store($E$, $F$)
            \State \Return Handoff($F$, $Q$)
        \Else
            \State \Return Fail(``No capable agent found.'')
        \EndIf
    \Else
        \State SemanticCache.Store($E$, $Best$)
        \State \Return Handoff($Best$, $Q$)
    \EndIf
\EndFunction
\end{algorithmic}
\end{algorithm}

\begin{algorithm}[!h]
\small
\caption{Agent-side retrieval flow}
\label{alg:agent-side-retrieval-request}
\begin{algorithmic}[1]
\Function{HandleProbeRequest}{Query $Q$, Requestor $R$}
    \If{\textbf{not} IsAuthorized($R$)}
        \State \Return ``Not Authorized''
    \EndIf
    \State $E \gets$ Embed($Q$)
    \State $S \gets$ AgentCache.FindSimilar($E$)
    \If{$S \neq \varnothing$}
        \State \Return ``OK''
    \EndIf
    \State $D \gets$ KnowledgeBase.Search($Q$)
    \If{$D = \varnothing$}
        \State AgentCache.Store($E$, ``KO'')
        \State \Return ``KO''
    \EndIf
    \If{LLM.AnalyzeCompleteness($Q$, $D$)}
        \State AgentCache.Store($E$, ``OK'')
        \State \Return ``OK''
    \Else
        \State AgentCache.Store($E$, ``Partial Content'')
        \State \Return ``Partial Content''
    \EndIf
\EndFunction
\end{algorithmic}
\end{algorithm}

\section{Experiments}
To validate our Knowledge Base-Aware (KBA) orchestration, we performed a series of controlled experiments comparing its performance against a standard description-driven baseline. We developed a custom benchmark suite to simulate a dynamic multi-agent environment where routing decisions are non-trivial and agent knowledge can evolve.

This section details our evaluation framework and presents the results. We begin by describing the benchmark architecture and the simulated task environment. We then define the performance metrics used for the comparison, focusing on routing accuracy and system latency. Finally, we present and analyze the empirical results of our KBA approach against the baseline.

\subsection{Setup}
To provide a robust comparison, we designed an experimental setup consisting of a simulated multi-agent environment, the two orchestration systems under evaluation (description-driven and the Knowledge Base-Aware one), and a set of controlled variables. We selected Google’s Agent Development Kit (ADK)~\cite{googleADKMultiAgents} as the core framework for the evaluation setup since it simplifies development and provides a neutral, state-of-the-art implementation of a description-driven orchestrator to serve as our baseline.

\subsubsection{Agents and Knowledge Bases}
We implemented seven domain-specific agents, each representing a typical corporate function:

\begin{itemize}
\item \textbf{Accounting}: Financial operations, workflows, compliance, and transactional queries (Coupa, Concur, NetSuite, SAP).
\item \textbf{HR}: HR processes, benefits, time-off, compliance, and employee experience (Workday, SAP SuccessFactors, ServiceNow).
\item \textbf{IT}: IT security, device compliance, identity management, and policy enforcement.
\item \textbf{Legal}: Legal guidance, contracts, IP, compliance, and risk.
\item \textbf{Marketing}: Brand governance, campaigns, content workflows (Figma, Marketo, Google Ads).
\item \textbf{R\&D}: Product lifecycle, code quality, hardware, innovation, and documentation.
\item \textbf{Sales}: Sales operations, pipeline, enablement, pricing exceptions (Salesforce, Gainsight, CPQ, Xactly).
\end{itemize}

Each agent was equipped with a dedicated knowledge base, created ad hoc using generative AI tools and structured as a set of domain-specific policies with the enterprise-wide assistant.

To evaluate routing performance, we constructed a test set of 140 questions, with 20 questions specifically designed to be answerable by each of the seven agents. This design ensures that a correct answer is always available within the system, allowing the evaluation to focus strictly on the orchestrator's ability to route the query correctly.

\subsection{System Under eEaluation}
Within the benchmark environment, we deployed and tested two distinct orchestrators: \textit{description-driven Orchestrator} as a baseline and \textit{our proposed KBA Orchestrator}.

\subsubsection{Baseline: Description-driven orchestrator}
Our baseline is a standard description-driven orchestrator, implemented using the ADK's \textit{LlmAgent} with \emph{LLM-Driven Delegation}~\cite{googleADKAgentTransfer}. This system’s routing logic relies solely on matching the user's query against the static, textual descriptions of the agents.

\subsubsection{Proposed: KBA Orchestrator}
Our proposed KBA (Knowledge Base-Aware) orchestrator was implemented by extending the baseline system. We integrated our custom components for \textit{Confidence-Based Initial Routing} and \textit{Dynamic Knowledge Probing} to augment the static description matching, as defined in Section~\ref{section:methodology}.

\subsection{Experimental Variables}
To conduct a thorough comparison, we systematically tested the performance of both orchestrators across several key variables.

\paragraph{Agent Description Variants}
As routing quality is highly dependent on the specific wording of agent descriptions (see Section~\ref{sec:agent-orchestration}), we systematically varied these along two dimensions:

\begin{itemize}
\item \textbf{Length}: Three variants were prepared: \emph{Basic} (120–125 characters), \emph{Balanced} (400–500 characters), and \emph{Detailed} (approximately 1000 characters).
\item \textbf{Content source}: Two sources were considered: \emph{Generic} descriptions derived from high-level functional requirements, and \emph{Fine-tuned} descriptions created by summarizing each agent's actual knowledge base.
\end{itemize}

This procedure resulted in six distinct description variants ($3 \times 2$) for each of the seven agents. An illustrative example of these variants is provided in Appendix~\ref{app:sub_agent_desc}.

\paragraph{Model temperature}
For both the baseline and KBA orchestrators, we experimented with multiple LLM temperature settings to analyze the impact of model creativity versus determinism on routing accuracy.

\subsection{Results and Analysis}

This section presents a comprehensive evaluation of our KBA orchestration framework against the standard description-driven baseline. The analysis is structured as follows: we first determine the optimal parameters for both systems, then conduct a comparative analysis of their classification performance and operational costs.

\subsubsection{Parameter Tuning and Baseline Configuration}

To ensure a fair comparison, we first identified the optimal configuration for each orchestrator by evaluating the impact of model temperature and agent description quality.

\paragraph{Model temperature}
We evaluated three temperature settings (0.2, 0.5, and 0.8) to assess the impact of sampling stochasticity on routing performance. As summarized in Table~\ref{tab:temp_tuning}, varying the temperature in either direction produced only minor and inconsistent changes in accuracy across both orchestrators. Since no systematic improvements were observed, we fixed the temperature at \textbf{0.2} for all subsequent experiments in order to minimize randomness and ensure deterministic, reproducible outcomes.

\begin{table}[h!]
\centering
\begin{tabular}{llc}
\toprule
\textbf{Orchestrator} & \textbf{Temperature} & \textbf{Accuracy $\Delta$ (vs. 0.2)} \\
\midrule
\multirow{2}{*}{Description-driven}
& 0.50 & +0.7\% \\
& 0.80 & -2.1\% \\
\midrule
\multirow{2}{*}{KBA}
& 0.50 & +1.4\% \\
& 0.80 & +0.7\% \\
\bottomrule
\end{tabular}
\caption{Routing accuracy variation across temperature settings relative to 0.2.}
\label{tab:temp_tuning}
\end{table}

\paragraph{Agent cards}
We investigated how the quality of agent descriptions affects routing by evaluating six variants formed by crossing three lengths (Basic, Balanced, Detailed) with two content sources (Generic, Fine-tuned). 

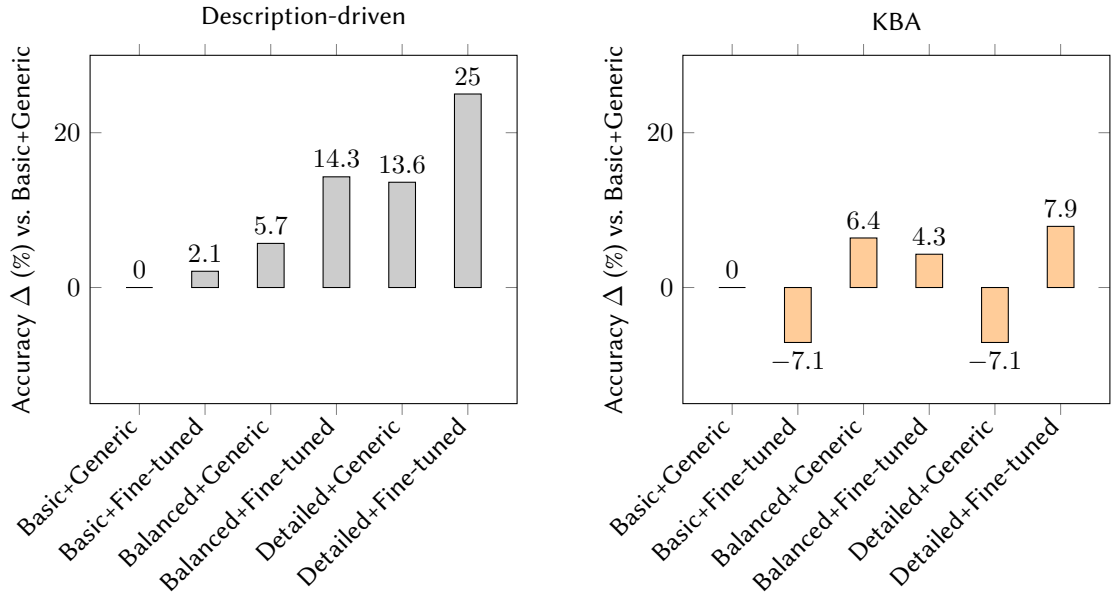
\begin{figure}[h!]
    \centering
    \begin{minipage}[t]{0.48\textwidth}
    \centering
    \begin{tikzpicture}
      \begin{axis}[
        ybar,
        width=\textwidth,
        height=6.2cm,
        bar width=10pt,
        enlarge x limits=0.15,
        ylabel={Accuracy $\Delta$ (\%) vs.\ Basic+Generic},
        symbolic x coords={Basic+Generic,Basic+Fine-tuned,Balanced+Generic,Balanced+Fine-tuned,Detailed+Generic,Detailed+Fine-tuned},
        xtick=data,
        nodes near coords,
        nodes near coords align={vertical},
        ymin=-15, ymax=30,
        xticklabel style={rotate=45, anchor=east},
        title={Description-driven}
      ]
        \addplot[fill=gray!40, draw=black] coordinates {
          (Basic+Generic,0.0)
          (Basic+Fine-tuned,2.1)
          (Balanced+Generic,5.7)
          (Balanced+Fine-tuned,14.3)
          (Detailed+Generic,13.6)
          (Detailed+Fine-tuned,25.0)
        };
      \end{axis}
    \end{tikzpicture}
    \end{minipage}
    \hfill
    \begin{minipage}[t]{0.48\textwidth}
    \centering
    \begin{tikzpicture}
      \begin{axis}[
        ybar,
        width=\textwidth,
        height=6.2cm,
        bar width=10pt,
        enlarge x limits=0.15,
        ylabel={Accuracy $\Delta$ (\%) vs.\ Basic+Generic},
        symbolic x coords={Basic+Generic,Basic+Fine-tuned,Balanced+Generic,Balanced+Fine-tuned,Detailed+Generic,Detailed+Fine-tuned},
        xtick=data,
        nodes near coords,
        nodes near coords align={vertical},
        ymin=-15, ymax=30,
        xticklabel style={rotate=45, anchor=east},
        title={KBA}
      ]
        \addplot[fill=orange!40, draw=black] coordinates {
          (Basic+Generic,0.0)
          (Basic+Fine-tuned,-7.1)
          (Balanced+Generic,6.4)
          (Balanced+Fine-tuned,4.3)
          (Detailed+Generic,-7.1)
          (Detailed+Fine-tuned,7.9)
        };
      \end{axis}
    \end{tikzpicture}
    \end{minipage}

    \caption{Accuracy change relative to the \emph{Basic + Generic} description baseline. \textbf{Left}: description-driven orchestrator shows strong dependence on description quality. \textbf{Right}: KBA orchestrator is more robust, though its highest accuracy is also achieved with \emph{Detailed + Fine-tuned}.}
    \label{fig:description_impact}
\end{figure}

The results, visualized in Figure~\ref{fig:description_impact}, highlight a key difference between the two orchestration strategies. The description-driven orchestrator is \textbf{highly sensitive} to description quality: accuracy steadily increases with richer, fine-tuned descriptions, reaching gains of up to \(+25\%\) for \emph{Detailed + Fine-tuned}. In contrast, the KBA orchestrator is \textbf{largely insensitive}, with only modest fluctuations across variants. While its best accuracy is also achieved under the \emph{Detailed + Fine-tuned} setting, its performance remains consistently high even with much simpler descriptions. This relative robustness reduces the engineering effort required to design and maintain agent descriptions in large multi-agent systems.

\paragraph{Configurations for final comparison}
Guided by the preceding analysis, we selected two representative configurations for each orchestrator in order to compare their performance under both minimal and optimized description conditions:
\begin{itemize}
    \item \textbf{Baseline-Low}: description-driven orchestrator with \emph{Basic + Generic} descriptions, representing the lowest-effort setup.
    \item \textbf{Baseline-High}: description-driven orchestrator with \emph{Detailed + Fine-tuned} descriptions, corresponding to its empirically optimal configuration.
    \item \textbf{KBA-Low}: KBA orchestrator with \emph{Basic + Generic} descriptions, testing performance under minimal descriptive information.
    \item \textbf{KBA-High}: KBA orchestrator with \emph{Detailed + Fine-tuned} descriptions, reflecting its highest observed accuracy.
\end{itemize}

\subsubsection{Comparative Performance Analysis}

We evaluated the four configurations on our 140-question test set, analyzing both \emph{classification quality} and \emph{operational cost}.

\paragraph{Routing accuracy and classification performance}
\textbf{KBA-Low} reaches \textbf{87.1\%} accuracy, an absolute gain of 43.5 points over \textbf{Baseline-Low} (43.6\%) and +18.5 points over \textbf{Baseline-High} (68.6\%). The best result is obtained by \textbf{KBA-High} at \textbf{95.0\%}. The same pattern holds for weighted precision, recall, and F1-score.
Table~\ref{tab:inline_metrics} presents the classification metrics.

\newlength{\metricbarwidth}
\setlength{\metricbarwidth}{2.8cm}
\newcommand{\metricbar}[2]{%
\begin{tikzpicture}[baseline=0.3ex]
  \draw[black!25] (0,0) rectangle (\metricbarwidth,0.16cm);
  \fill[#2] (0,0) rectangle ({#1*\metricbarwidth},0.16cm);
\end{tikzpicture}%
}

\begin{table}[h!]
\centering
\small
\begin{tabular}{lcc}
\toprule
\textbf{Metric} & \textbf{Baseline Configurations} & \textbf{KBA Configurations} \\
\midrule
Accuracy &
\begin{tabular}{@{}l@{}}
Low: 43.6\% \; \metricbar{0.436}{gray!45}\\
High: 68.6\% \; \metricbar{0.686}{gray!70}
\end{tabular} &
\begin{tabular}{@{}l@{}}
Low: 87.1\% \; \metricbar{0.871}{orange!55}\\
High: 95.0\% \; \metricbar{0.950}{orange!85}
\end{tabular} \\
\midrule
Precision (weighted) &
\begin{tabular}{@{}l@{}}
Low: 58.0\% \; \metricbar{0.580}{gray!45}\\
High: 76.3\% \; \metricbar{0.763}{gray!70}
\end{tabular} &
\begin{tabular}{@{}l@{}}
Low: 91.3\% \; \metricbar{0.913}{orange!55}\\
High: 95.4\% \; \metricbar{0.954}{orange!85}
\end{tabular} \\
\midrule
F1-Score (weighted) &
\begin{tabular}{@{}l@{}}
Low: 40.9\% \; \metricbar{0.409}{gray!45}\\
High: 70.0\% \; \metricbar{0.700}{gray!70}
\end{tabular} &
\begin{tabular}{@{}l@{}}
Low: 88.9\% \; \metricbar{0.889}{orange!55}\\
High: 95.0\% \; \metricbar{0.950}{orange!85}
\end{tabular} \\
\bottomrule
\end{tabular}
\caption{Classification performance. Each cell shows both Low and High configurations for the respective orchestrator with inline bars scaled to 100\%. Given single-label routing with exactly one correct agent per query, recall equals accuracy; we therefore omit a separate recall column}
\label{tab:inline_metrics}
\end{table}

\paragraph{Operational costs: latency and token usage}
The improved performance of KBA is associated with increased resource consumption (Table~\ref{tab:cost_metrics}). The dynamic knowledge probing step introduces higher token usage and longer execution times compared to the description-driven baseline. However, this overhead primarily reflects the absence of caching in our experimental setup. As discussed in Section~\ref{section:methodology}, the introduction of a semantic cache is expected to substantially mitigate these costs in practice, especially under workloads with recurring queries.

\begin{table}[h]
\centering
\small
\begin{tabular}{lcccc}
\toprule
\textbf{Metric} & \textbf{Baseline-Low} & \textbf{Baseline-High} & \textbf{KBA-Low} & \textbf{KBA-High} \\
\midrule
Input Tokens        & 103{,}022 & 309{,}382 & 240{,}089 & 659{,}069 \\
Output Tokens       & 5{,}339   & 3{,}108   & 15{,}299  & 16{,}540 \\
Execution Time (s)  & 106       & 97        & 567       & 577 \\
\bottomrule
\end{tabular}
\caption{Resource usage without caching optimization. KBA’s dynamic probing yields higher token consumption and longer execution times.}
\label{tab:cost_metrics}
\end{table}

\section{Discussion}

The experimental results demonstrate that Knowledge Base-Aware (KBA) Orchestration provides a substantial improvement over conventional description-driven approaches, particularly in environments where agent descriptions are incomplete, ambiguous, or overlapping. Several insights can be drawn from these findings.

\subsection{Impact of Dynamic Probing and Semantic Caching}
The introduction of dynamic knowledge probing proved to be the decisive factor in improving routing accuracy. Unlike static description matching, which relies on human-authored summaries that can rapidly become outdated, KBA leverages the agents’ internal knowledge bases to provide context-sensitive relevance signals at runtime. This shift effectively transforms orchestration from a one-shot classification task into an adaptive dialogue between orchestrator and agents. The semantic cache further enhances efficiency, amortizing the probing cost across recurring queries and reducing the performance penalty of the probing mechanism over time.

\subsection{Reduced Sensitivity to Description Quality}
A key practical advantage of KBA orchestration lies in its robustness to the quality of agent descriptions. While the description-driven baseline required carefully curated, fine-tuned descriptions to achieve acceptable accuracy, KBA maintained strong performance even with minimal descriptions. This lowers the engineering burden of system deployment and maintenance, making orchestration viable at larger scales where continuously updating agent profiles is impractical.

\subsection{Trade-offs: Accuracy versus Resource Consumption}
The improved performance of KBA comes at the cost of increased token usage and latency due to dynamic probing. Although these overheads are significant in cold-start scenarios, they are mitigated by caching and by the fact that probing is only invoked under conditions of uncertainty. From a system design perspective, this introduces a tunable trade-off: organizations may prioritize accuracy in high-stakes domains (e.g., legal, compliance, healthcare) or optimize for efficiency in low-stakes or high-throughput environments. Importantly, the probing mechanism is privacy-preserving by design, ensuring that agents expose only lightweight signals rather than sensitive data.

\subsection{Limitations and Future Directions}
Despite its advantages, KBA orchestration faces several challenges. First, probing incurs synchronization costs when scaling to very large agent pools, potentially creating latency bottlenecks. Second, we did not define standardized patterns for implementing the \textit{OK/KO} signaling mechanism on the agent side during Dynamic Knowledge Probing. While leaving this freedom to implementers can be seen as a strength, allowing integration with diverse architectures, it also introduces variability in performance. Providing implementation guidelines or best practices would help ensure consistent efficiency across deployments. Third, the design and use of the semantic cache may prove tricky in practice: cache invalidation, expiration policies, and handling of semantically similar but contextually distinct queries require careful consideration to avoid erroneous routing decisions. Finally, our evaluation assumed cooperative agents; adversarial or misconfigured agents could misreport capabilities, raising trust and governance concerns. 

These issues highlight that while KBA orchestration is effective in principle, its practical implementation raises several open challenges. Future research should focus on refining the OK/KO signaling process with performance-oriented guidelines, improving strategies for managing semantic caches, reducing synchronization costs in large agent pools, and strengthening resilience against misconfigured or adversarial agents.

\section{Conclusion}

This paper introduced Knowledge Base-Aware (KBA) Orchestration, a novel approach for improving task routing in multi-agent systems. By augmenting static agent descriptions with dynamic, privacy-preserving relevance signals derived from agents’ private knowledge bases, KBA overcomes key limitations of description-driven orchestration. Our empirical evaluation shows that KBA achieves significantly higher routing accuracy and robustness to description quality, while maintaining adaptability to evolving agent expertise.

Although KBA incurs higher computational overhead, semantic caching and selective probing provide effective mechanisms to mitigate these costs. The result is a more adaptive and resilient orchestration strategy that balances efficiency, privacy, and accuracy in dynamic environments.

In conclusion, KBA orchestration represents a meaningful step toward scalable, reliable, and knowledge-aware multi-agent systems. Future research should focus on refining probing mechanisms, standardizing signaling practices, and advancing cache management strategies to ensure robustness in large-scale deployments. Collectively, these directions point toward a new generation of orchestration frameworks capable of unlocking the full potential of multi-agent intelligence in dynamic and diverse environments.

\newpage
\bibliography{citations}

\newpage
\appendix

\section{Experimental Setup Resources Examples}
\subsection{Subagent Knowledge Base}

We report below two representative examples of policies provided to subagents. The first pertains to the HR agent and concerns remote work eligibility; the second refers to the IT agent and details the procedure for enrolling in multi-factor authentication (MFA).

\subsubsection{HR Agent Policy}
\begin{mdframed}[linewidth=1pt]
\textbf{Maternity and Paternity Leave Policy and Procedure}

\textbf{1. Purpose} \\
This policy outlines the eligibility criteria, entitlements, and the application process for maternity and paternity leave within the organization. The objective is to support employees during the significant life event of welcoming a child, ensuring compliance with applicable labor laws and promoting work-life balance.

\textbf{2. Scope} \\
This policy applies to all full-time and part-time employees who meet the eligibility criteria set forth herein.

\textbf{3. Eligibility}
\begin{itemize}
\item \textbf{Maternity Leave:} All female employees who have completed at least 12 months of continuous service with the organization prior to the expected date of childbirth.
\item \textbf{Paternity Leave:} All male employees who have completed at least 12 months of continuous service with the organization prior to the expected date of childbirth or adoption.
\end{itemize}

\textbf{4. Entitlement}
\begin{itemize}
\item \textbf{Maternity Leave:} Up to 16 weeks, potentially including paid and unpaid portions in line with statutory regulations and company policy.
\item \textbf{Paternity Leave:} Up to 2 weeks, which may be taken consecutively or flexibly within 12 weeks of birth or adoption.
\end{itemize}

\textbf{5. Leave Application Procedure}

\textbf{5.1 Notification} \\
Employees must inform their supervisor and HR in writing at least 8 weeks in advance of the intended leave start date (or as early as practicable in unforeseen situations).

\textbf{5.2 Application via SAP SuccessFactors}
\begin{enumerate}
\item \textbf{Login:} Access SAP SuccessFactors through the company intranet or portal.
\item \textbf{Navigate to Time Off:} Open the ‘Time Off’ module.
\item \textbf{Request Leave:} Select the relevant leave type (Maternity or Paternity).
\item \textbf{Specify Dates:} Enter intended start and end dates.
\item \textbf{Attach Documentation:} Upload necessary medical/adoption documents.
\item \textbf{Submit:} Confirm details and send for approval.
\end{enumerate}

\textbf{5.3 Approval Process}
\begin{itemize}
\item The supervisor reviews and approves the initial request.
\item HR provides final validation.
\item Employees are notified via SAP SuccessFactors within 5 business days.
\end{itemize}

\textbf{6. Return to Work} \\
Employees returning early or requesting extensions must update their leave request and inform HR. A medical certificate may be required post-maternity leave.

\textbf{7. Record Keeping and Confidentiality} \\
All related documentation is treated as confidential under the company’s data protection policy.

\textbf{8. Policy Review} \\
Reviewed annually or as needed to maintain legal and procedural compliance.

\textbf{References}
\begin{itemize}
\item Applicable labor laws on family leave
\item SAP SuccessFactors Employee Central User Guide
\item Company Data Protection Policy
\end{itemize}

\textit{For questions or assistance with the process, contact the HR department.}
\end{mdframed}

\vspace{1.5em}

\subsubsection*{IT Agent Policy}
\begin{mdframed}[linewidth=1pt]
\textbf{Company Policy and Procedure Document} \\
\textbf{Multi-Factor Authentication (MFA) Enrollment Using Microsoft Authenticator}
\textbf{Scope:} Applies to all users accessing MFA-protected systems.

\textbf{1. Purpose} \\
Defines the procedure for enrolling in MFA using Microsoft Authenticator to secure company resources.

\textbf{2. Policy Statement} \\
All users must enroll in MFA before accessing protected resources.

\textbf{3. Scope} \\
Applies to employees, contractors, and systems including Microsoft 365, Azure AD, VPN, and internal portals.

\textbf{4. Roles and Responsibilities} \\
\begin{itemize}
\item \textbf{IT Security:} Configure MFA and support users.
\item \textbf{End Users:} Enroll promptly and maintain access.
\item \textbf{Managers:} Ensure compliance within teams.
\end{itemize}

\textbf{5. Definitions} \\
\begin{itemize}
\item \textbf{MFA:} Two or more verification methods.
\item \textbf{Microsoft Authenticator:} Mobile-based MFA app.
\item \textbf{Azure AD:} Identity and access management service.
\end{itemize}

\textbf{6. Prerequisites} \\
\begin{itemize}
\item Compatible mobile device.
\item Email access and Azure AD credentials.
\end{itemize}

\textbf{7. Enrollment Procedure}
\begin{enumerate}
\item Install the app from App Store/Play Store.
\item Log into \url{https://portal.office.com} and follow MFA prompts.
\item Configure app as authentication method.
\item Scan QR code to link device.
\item Approve test notification.
\item Finalize enrollment.
\item Enable cloud backup and set recovery options.
\end{enumerate}

\textbf{8. Usage Guidelines} \\
\begin{itemize}
\item Approve only self-initiated MFA prompts.
\item Report suspicious activity immediately.
\item Keep the app and mobile device secure.
\end{itemize}

\textbf{9. Support and Escalation} \\
Contact IT Helpdesk for assistance or reset requests.

\textbf{10. Enforcement} \\
Non-compliance may lead to access suspension. Periodic audits will be conducted.

\textbf{11. References} \\
\begin{itemize}
\item \url{https://aka.ms/authapp}
\item \url{https://portal.office.com}
\item \url{https://docs.microsoft.com/azure/active-directory/authentication/howto-mfa}
\end{itemize}
\end{mdframed}

\subsection{Subagent Descriptions}
\label{app:sub_agent_desc}
As explained earlier, each subagent is provided with a textual description that varies in both length and level of detail. Descriptions also differ in their degree of procedural specificity, which we categorize as either \emph{Generic} or \emph{Fine-tuned}.

Below we provide illustrative examples for the HR subagent:

\paragraph{HR Agent - Basic Description, Generic (Not Fine-tuned)}
\begin{quote}
Deals with employee-related topics like policies, benefits, and development.
\end{quote}

\paragraph{HR Agent - Basic Description, Fine-tuned}
\begin{quote}
Handles HR, benefits, time-off, compliance, and employee experience processes across platforms like Workday, SAP SuccessFactors, and ServiceNow.
\end{quote}

\paragraph{HR Agent - Detailed Description, Generic (Not Fine-tuned)}
\begin{quote}
Responds to inquiries regarding vacation and leave policies;  
health insurance and retirement benefits;  
grievance procedures;  
performance review processes;  
remote work policies;  
employee data management;  
training and development opportunities;  
workplace dress code and company values;  
salary and compensation structures;  
sick leave regulations;  
internal job postings and transfers;  
employee assistance programs;  
diversity and inclusion initiatives;  
expense reporting;  
onboarding and offboarding procedures;  
employee conduct and discipline;  
career advancement opportunities;  
payroll inquiries;  
social media usage at work;  
employee feedback mechanisms;  
mental health resources;  
employee well-being initiatives;  
sabbatical and flexible working arrangements;  
workplace relationships;  
employee referrals;  
learning and development strategies;  
workplace accidents and disability accommodations;  
political activities at work;  
employment records access;  
commuter benefits;  
performance improvement plans;  
discrimination complaints;  
employment contracts.
\end{quote}

\paragraph{HR Agent - Detailed Description, Fine-tuned}
\begin{quote}
This subagent is specialized in end-to-end HR operations, workplace compliance, employee experience programs, and digital workforce systems. It supports employees in navigating and resolving queries across topics such as maternity/paternity leave, PTO requests, ergonomic compliance, performance reviews, D\&I training, referral programs, I-9/E-Verify verification, exit processes, benefit enrollment via Mercer Marketplace, payroll corrections, job applications, grievances, accommodations, and learning \& development. It provides step-by-step guidance on using tools like Workday, SAP SuccessFactors, Ceridian Dayforce, Kronos, Cornerstone LMS, Zendesk HR, and ServiceNow, ensuring timely submissions, proper documentation, and policy adherence. It also manages workflows related to wellness reimbursements, job shadowing, background checks (e.g., via Checkr), GDPR and CCPA data requests, and internal mobility. Designed to support both employees and HR teams, it ensures smooth process execution, compliance with deadlines, and employee well-being throughout the employment lifecycle.
\end{quote}

\subsection{Questions test set}

The test set consists of 140 questions, evenly distributed across seven distinct subagents, with 20 questions per agent. Each question was derived from the respective agent's knowledge base to emulate realistic user queries directed at an orchestrator agent operating in a multi-agent environment.
Table~\ref{tab:testset_examples} presents a selection of representative queries from the test set, along with the corresponding agent responsible for answering each question.

\begin{table}[ht]
\centering
\renewcommand{\arraystretch}{1.2}
\begin{tabularx}{\textwidth}{>{\raggedright\arraybackslash}X l}
\toprule
\textbf{Example Question} & \textbf{Agent} \\
\midrule
How do I execute the force majeure addendum after approval? & \texttt{legal\_agent} \\
How do I report a gift or hospitality I received that is above \$50? & \texttt{legal\_agent} \\
What information must be included in a Written Warning and who needs to be present? & \texttt{hr\_agent} \\
What types of PTO can I select when submitting a request in the MyVacation system? & \texttt{hr\_agent} \\
What information is required when adding a new license key to the 1Password vault? & \texttt{it\_agent} \\
What should I do if my email signature does not match the standardized format? & \texttt{it\_agent} \\
Who is responsible for implementing approved Chart of Accounts changes in SAP S/4HANA? & \texttt{accounting\_agent} \\
What should I do if my receipt cannot be captured clearly by the OCR and I need to submit it manually? & \texttt{accounting\_agent} \\
What should I check before submitting my post to the Approval Queue? & \texttt{marketing\_agent} \\
What are the responsibilities of the QA team in maintaining landing page accessibility? & \texttt{marketing\_agent} \\
What steps should I follow to handle prospects who unsubscribe or opt out of emails? & \texttt{sales\_agent} \\
What information and documentation do I need to provide when submitting a pricing exception request in Deal Desk Jira? & \texttt{sales\_agent} \\
How should I identify and classify PII before starting the anonymization process? & \texttt{research\_and\_development\_agent} \\
How do I properly submit a request for creating a new feature flag? & \texttt{research\_and\_development\_agent} \\
\bottomrule
\end{tabularx}
\caption{Representative examples from the test set and their assigned subagents.}
\label{tab:testset_examples}
\end{table}

\end{document}